\documentclass[aps,pra,twocolumn,superscriptaddress,groupedaddress]{revtex4}  % for review and submission
\usepackage{graphicx}  % needed for figures
\usepackage{dcolumn}   % needed for some tables
\usepackage{bm}        % for math
\usepackage{amssymb}   % for math
\usepackage{amsmath}   % for math
\usepackage{braket}

\begin{document}

% The following information is for internal review, please remove them for submission

% the following line is for submission, including submission to the arXiv!!
%\hspace{5.2in} \mbox{Fermilab-Pub-04/xxx-E}

\title{A general inseparability criterion for non-Gaussian states}
\author{Prasoon K.~Shandilya}
\thanks{shandilya.prasoon@gmail.com}
\affiliation{Indian Institute of Science Education and Research Kolkata, Mohanpur 741246, West Bengal, India}

\author{Prasanta K. ~Panigrahi}
\thanks{pprasanta@iiserkol.ac.in}
\affiliation{Indian Institute of Science Education and Research Kolkata, Mohanpur 741246, West Bengal, India}

\vskip 0.25cm
       % D0 authors (remove the first 3 lines
                             % of this file prior to submission, they
                             % contain a time stamp for the authorlist)
                             % (includes institutions and visitors)
\date{\today}

\begin{abstract}
%\iffalse
We have derived a general separability criterion for a class of two mode non-Gaussian continuous variable systems, obtained earlier using PPT, violation of which provides sufficient condition for entanglement. It has been obtained by utilizing the Cauchy-Schwarz inequality and from the basic definition of separable states. This criterion coincides with the work of Agarwal and Biswas \cite{d} which involved inequality involving higher order correlation, for testing entanglement in non-Gaussian states.
%\fi

\end{abstract}

\maketitle
\section{Introduction}

Peres-Horodecki criterion, also known as positive partial transpose (PPT), is a separable condition for a general bipartite state of dimensions 2$\times$2 (two-qubits) and 2$\times$3 (qubit-qutrit) \cite{a}. It can be used for both pure and mixed states. Peres proposed PPT as a necessary condition for every separable state, implying that every separable state must satisfy this condition and some entangled states may satisfy it as well. Subsequently, Horodecki showed that it was a necessary and sufficient condition for separable states of 2$\times$2 and 2$\times$3 dimensions \cite{c} and only sufficient for higher dimensional states. Peres-Horodecki criterion states that if the partial transpose of a bipartite density matrix has at least one negative eigenvalue, then the state must be entangled. The other well known partial trace criterion ($\rho_A=Tr_B\{\rho_{AB}\}$) is limited to only pure states as it loses all the information about the subsystem for the mixed state. The state is entangled if the reduced density matrix obtained by performing a partial trace represents a mixed state and is separable if the reduced density matrix remains a pure
state.

Recently, Simon has shown that partial transpose operation on the continuous variable bipartite density operator corresponds to a mirror reflection which inverts only the momentum of one particle \cite{b} in a two mode bipartite system described by annihilation operators \(\hat a = (\hat x_a + i\hat p_a)/\sqrt[]{2}\), \(\hat b = (\hat x_b+i\hat p_b)/\sqrt[]{2}\) and creation operators \(\hat a^\dagger = (\hat x_a - i\hat p_a)/\sqrt[]{2}\), \(\hat b^\dagger = (\hat x_b-i\hat p_b)/\sqrt[]{2}\). Evidently, under the partial transpose, $\hat a \rightarrow \hat a$, $\hat a^\dagger \rightarrow \hat a^\dagger$, $\hat b \rightarrow \hat b^\dagger$ and $\hat b^\dagger \rightarrow \hat b$. Using PPT, Simon successfully concluded that the Peres-Horodecki criterion is a necessary and sufficient condition for separability, for all bipartite Gaussian continuous variable states. These inequalities involved variances of relative position and total momentum coordinates of the two subsystems and thus can be verified experimentally \cite{g}.

Subsequently, Duan \textit{et al} \cite{l} proposed equivalent inequalities independently using the positivity of the quadratic forms. It proved that for all the Gaussian continuous variable states, violation of these inequalities provides a necessary and sufficient criterion for entanglement. Mancini \textit{et al} \cite{m} derived a different form of criterion for entanglement involving second-order moments. Separability criteria by Duan \textit{et al} and Mancini \textit{et al} are interrelated with each other. Agarwal and Biswas has shown  that Duan \textit{et al} inseparability criterion is derived generally and is a weaker inseparability criterion compared to Mancini \textit{et al} inseparability criterion for Gaussian states \cite{d}. 

Various criteria for entanglement involving second-order moments have been tested for the Gaussian states \cite{h,k}. However, the existing inseparability criteria for entangled non-Gaussian states based on second-order correlations do not provide correct information about the inseparability \cite{d}. These states admit, SU(2) and SU(1,1) algebraic structures making them ideal for taking higher order correlations. In general, SU(2) operators are related to the interactions between photon and atom, whereas SU(1,1) is related to the non-linear parametric generation and conversion of two photons \cite{i}.

In this paper, we investigate a class of non-Gaussian states, which appear in quantum optical systems and have been studied recently in connection with PPT \cite{d}. We derive a general inseparability criteria for a class of two mode non-Gaussian continuous variable system utilizing the Cauchy-Schwarz inequality and from the basic definition of separable states.This criterion coincides with the work of Agarwal and Biswas \cite{d} which involved inequality involving higher order correlation, for testing entanglement in non-Gaussian states.

\section{Separability criteria using PPT}
 We consider following set of operators satisfying SU(2) algebra
\begin{equation}
S_x= \frac{a^\dagger b + ab^\dagger}{2}, S_y= \frac{a^\dagger b - ab^\dagger}{2i}, S_z= \frac{a^\dagger a - b^\dagger b}{2}
\end{equation} The operator $S_i$ obey the algebra of angular momentum operators $[S_i,S_j] = i\epsilon _{ijk}S_k$ and leads us to the following uncertainty relation \cite{v}:
\begin{equation}
\Delta S_x \Delta S_y \geq \frac{1}{2}\mid \langle S_z \rangle \mid
\label{e}
\end{equation}
which leads to
\begin{equation}
\Delta \Bigg[\frac{a^\dagger b + ab^\dagger}{2}\Bigg] \Delta \Bigg[\frac{a^\dagger b - ab^\dagger}{2i}\Bigg] \geq \frac{1}{2}\Bigg| \Bigg\langle\frac{a^\dagger a - b^\dagger b}{2}\Bigg\rangle \Bigg|
\label{S_ineq}
\end{equation}

We now consider the operators of SU(1,1) algebra  that satisfy $[K_x,K_y] = -iK_z$, $[K_y,K_z] = iK_x$ and $[K_z,K_x] = iK_y$, where
\begin{equation}
K_x= \frac{a^\dagger b^\dagger + ab}{2}, K_y= \frac{a^\dagger b^\dagger - ab}{2i}, K_z= \frac{a^\dagger a + b^\dagger b +1}{2}
\end{equation}
From uncertainty relation it follows that
\begin{equation}
\Delta K_x \Delta K_y \geq \frac{1}{2} |\langle K_z \rangle |
\label{f}
\end{equation}
which leads to 
\begin{equation}
\Delta \Bigg[\frac{a^\dagger b^\dagger + ab}{2}\Bigg] \Delta \Bigg[\frac{a^\dagger b^\dagger - ab}{2i}\Bigg] \geq \frac{1}{2}\Bigg| \Bigg\langle\frac{a^\dagger a + b^\dagger b +1}{2}\Bigg\rangle \Bigg|
\label{K_ineq}
\end{equation}
 If a state is separable, the associated density matrix remains as a valid density operator under partial transposition \cite{a}.
Under the partial transpose, above inequality turns into 
\begin{eqnarray}
\Delta \Bigg[\frac{a^\dagger b + ab^\dagger}{2}\Bigg] \Delta \Bigg[\frac{a^\dagger b - ab^\dagger}{2i}\Bigg] & \geq \frac{1}{2}\Bigg| \Bigg\langle\frac{a^\dagger a + bb^\dagger +1}{2}\Bigg\rangle \Bigg| \\
& \geq \frac{1}{2}\Bigg| \Bigg\langle\frac{a^\dagger a + b^\dagger b +2}{2}\Bigg\rangle \Bigg| \\
& \geq \frac{1}{2}\Bigg| \Bigg\langle\frac{a^\dagger a + b^\dagger b }{2}\Bigg\rangle \Bigg|
\label{K_ineq}
\end{eqnarray}
On analyzing inequality (\ref{S_ineq}) and inequality (\ref{K_ineq}) we can conclude that inequality (\ref{K_ineq}) is much stronger than the inequality (\ref{S_ineq}). Agarwal and Biswas has shown that for a composite system state of bosonic particles which are part of an inseparable non-Gaussian state of the Bell form do not violate the inequality (\ref{S_ineq}) but do violate the inequality (\ref{K_ineq}) \cite{d}.\\

Explicitly introducing $S_x$, $S_y$ and $S_z$ in inequality (\ref{K_ineq}) leads to
\begin{equation}
\Delta S_x \Delta S_y \geq \frac{1}{2}\mid \langle S_z \rangle  + \langle b^\dagger b \rangle \mid
\label{a}
\end{equation}

Therefore we can conclude that this criterion can certainly detect entanglement for a broad class of non-Gaussian SU(2) minimum uncertainty states. 

Inequality (\ref{a}) can be written as
\begin{equation}
\Delta S_x \Delta S_y \geq \frac{1}{4} \mid \langle N_a \rangle  + \langle N_b \rangle \mid
\label{b}
\end{equation}
where $a^\dagger a = N_a$ and $b^\dagger b = N_b$.\\

\section{A General Criterion}
We start with the inequality
\begin{align}
(\Delta S_x - \Delta S_y)^2 & \geq 0 \label{d} \\
(\Delta S_x)^2 + (\Delta S_y)^2 & \geq 2\Delta S_x \Delta S_y \nonumber 
\end{align}
which is in general true. Explicit calculation of $(\Delta S_x)^2 + (\Delta S_y)^2$ leads to
\begin{align}
(\Delta S_x)^2 + (\Delta S_y)^2 &= \frac{1}{2} \big[\langle a^\dagger bab^\dagger\rangle+\langle ab^\dagger a^\dagger b\rangle \nonumber \\ 
&-2\langle a^\dagger b\rangle\langle ab^\dagger\rangle \big] \nonumber \\
&= \frac{1}{2}\big[\langle(a^\dagger a+1)b^\dagger b \rangle + \langle a^\dagger a(b^\dagger b+1)\rangle \nonumber \\
&-2\mid \langle ab^\dagger  \rangle \mid^2 \big]
\label{m}
\end{align}
If the states are separable,
\begin{equation}
|\langle ab^\dagger  \rangle |^2 = | \langle a\rangle\langle b^\dagger  \rangle |^2
\end{equation}
For the separable states Eq. (\ref{m}) translates into
\begin{align}
(\Delta S_x)^2 + (\Delta S_y)^2 &= \frac{1}{2}[\langle(a^\dagger a+1) \rangle\langle b^\dagger b \rangle + \langle a^\dagger a \rangle\langle (b^\dagger b+1)\rangle \nonumber \\
&-2\mid \langle a\rangle\langle b^\dagger  \rangle \mid^2]\\
&= \frac{1}{2}[\langle(N_a+1) \rangle\langle N_b \rangle + \langle N_a \rangle\langle (N_b+1)\rangle \nonumber \\
&-2\mid \langle a\rangle\langle b^\dagger  \rangle \mid^2]
\label{l}
\end{align}
We note that the Cauchy-Schwarz inequality yields $| \langle a \rangle |^2 \leq \langle N_a \rangle $ and $| \langle b \rangle |^2 \leq \langle N_b \rangle $ implying
\begin{equation}
(\Delta S_x)^2 + (\Delta S_y)^2 \geq \frac{1}{2} \mid \langle N_a \rangle  + \langle N_b \rangle |
\label{c}
\end{equation}
Interestingly this inequality is same as the inequality (\ref{b}) which has been deduced using partial transpose. It can be shown by using inequality (\ref{b}) and inequality (\ref{d}), that is 
\begin{align}
(\Delta S_x)^2 + (\Delta S_y)^2 \geq 2\Delta S_x \Delta S_y\nonumber \textrm{ and}\\ 2\Delta S_x \Delta S_y \geq \frac{1}{2} \mid \langle N_a \rangle  + \langle N_b \rangle | \nonumber \\
\end{align}
This leads to
\begin{align}
(\Delta S_x)^2 + (\Delta S_y)^2 \geq \frac{1}{2} \mid \langle N_a \rangle  + \langle N_b \rangle |
\end{align}

In analogy to the method applied for finding entanglement condition for the operators satisfying SU(2) algebra, it is possible to find similar relations for SU(1,1) operators. Under partial transpose, the uncertainty relation for operators following SU(2) algebraic structure leads us to
\begin{align}
\Delta K_x \Delta K_y \geq \frac{1}{4} |\langle a^\dagger a - bb^\dagger \rangle | \nonumber \\
\geq \frac{1}{2}| \langle K_z -b^\dagger b -1\rangle |
\end{align}
This inequality violates the basic uncertainty relation (Inequality \ref{f}) which is valid for every physical state showing the inadequacy of using PPT for the entanglement criteria. Explicitly calculating the uncertainties of these variables and adding them also do not leads us to any new information other than the uncertainty relation.\\

\section{Conclusion}
In conclusion, we have derived a general separability inequality for a class of two mode non-Gaussian continuous variable systems, obtained earlier using PPT, violation of which provides sufficient condition for entanglement. It has been obtained by utilizing the Cauchy-Schwarz inequality and from the basic definition of separable states. We have also shown how PPT violates uncertainty relation for the operators satisfying SU(1,1) algebraic structure which makes it inadequate for general use for determining the entanglement criteria. Since the separability inequality involves expectation values of observables, they can be tested experimentally. In future, we hope to generalize such inseparability inequalities for other algebraic structures and multi-partite quantum systems. We also hope that our work can be extended to arrive at the PPT as an inherent geometric criterion of separability for the continuous variable states \cite{o} and to generate stronger condition for inseparability using entanglement witness operator \cite{n}.

\end{document}